\documentclass{eas}
\usepackage{graphicx}
\usepackage{url}
\usepackage{siunitx}

\def\etalsp {\em{et al.}\/}

\def\mag{\,{\rm mag}} 
\def\kpc{\,{\rm kpc}} 
\def\kms{\,{\rm km}/{\rm s}}
\def\deg{\hbox{$\null^\circ$}}

\urldef{\gaiaperformance}\url{http://www.cosmos.esa.int/web/gaia/science-performance#astrometric%20performance}

\begin{document}
\title{What can Gaia proper motions tell us about Milky Way dwarf galaxies?}
\runningtitle{Proper motions of dwarf galaxies from Gaia}
\author{Shoko Jin}
\author{Amina Helmi}
\author{Maarten Breddels$\dagger$}
\thanks{$\dagger$Kapteyn Astronomical Institute, University of Groningen, P.O. Box 800, 9700 AV Groningen, The Netherlands}

\begin{abstract}
We present a proper-motion study on models of the dwarf spheroidal galaxy Sculptor, based on the predicted proper-motion accuracy of Gaia measurements. Gaia will measure proper motions of several hundreds of stars for a Sculptor-like system. Even with an uncertainty on the proper motion of order 1.5 times the size of an individual proper-motion value of $\sim10\,\rm{mas/century}$, we find that it is possible to recover  Sculptor's systemic proper motion at its distance of $79\kpc$.
\end{abstract}

\maketitle

\section{Introduction}

Proper motions from Gaia will open a new window in the field of dynamical studies of Galactic substructure, eliminating the need for kinematic inferences in their absence. Access to 6D phase-space coordinates ({\it{i.e.}} knowledge of position and velocity vectors) for a vast number of Milky-Way stars will pave the way for greatly improved modelling of all components of the Milky Way, including its satellite galaxies and stellar streams, in turn placing stricter constraints on global properties of our Galaxy such as its total mass. This, however, relies on our ability to deduce robust proper motions of the substructures of interest.

We currently expect that Gaia will deliver proper motions accurate to \SI{330}{\micro as/yr} or better for solar-type (G2V) stars brighter than $G\sim20\,\mag$.\footnote{\gaiaperformance} This translates to a proper-motion accuracy of better than $1\kms$ out to a distance of $20\kpc$. Although Gaia is unprecedented in both the number of stars that it will observe for astrometry and the accuracy to which this will be performed, there are entire populations of stars --- namely dwarf galaxies residing in the outskirts of the Galaxy ---  for which Gaia's astrometric accuracy would, na\"ively speaking, not be sufficient to deduce reliable proper motions. This statement is based on the simple fact that the satellite galaxies are mostly at distances of several tens of kpc or even further, where the magnitude of the proper-motion uncertainties are expected to be of the order of, or even greater than, the actual stellar velocities to be measured.

In principle, Milky Way dwarf galaxies contain hundreds of stars down to the magnitude limit for proper-motion measurements from Gaia. The question posed here is whether we can benefit from these individual proper-motion measurements for stars belonging to dwarf galaxies at all and, if so, to what extent.

For this exercise, we use Sculptor as a testbed. At a distance of $79\kpc$ (Mateo \cite{1998ARA&A..36..435M}), its near-polar location at $(l,b) = (287.5,-83.2)\deg$ minimises Galactic contamination in its field, making this dwarf galaxy an ideal system for testing whether proper-motion data from Gaia will provide any insights into its global or internal dynamics. The following section describes the data used to set up our mock version of Sculptor and our analysis of the resulting system.

\section{Analysis of a mock Sculptor}

\begin{figure}
\includegraphics[width=12.4cm]{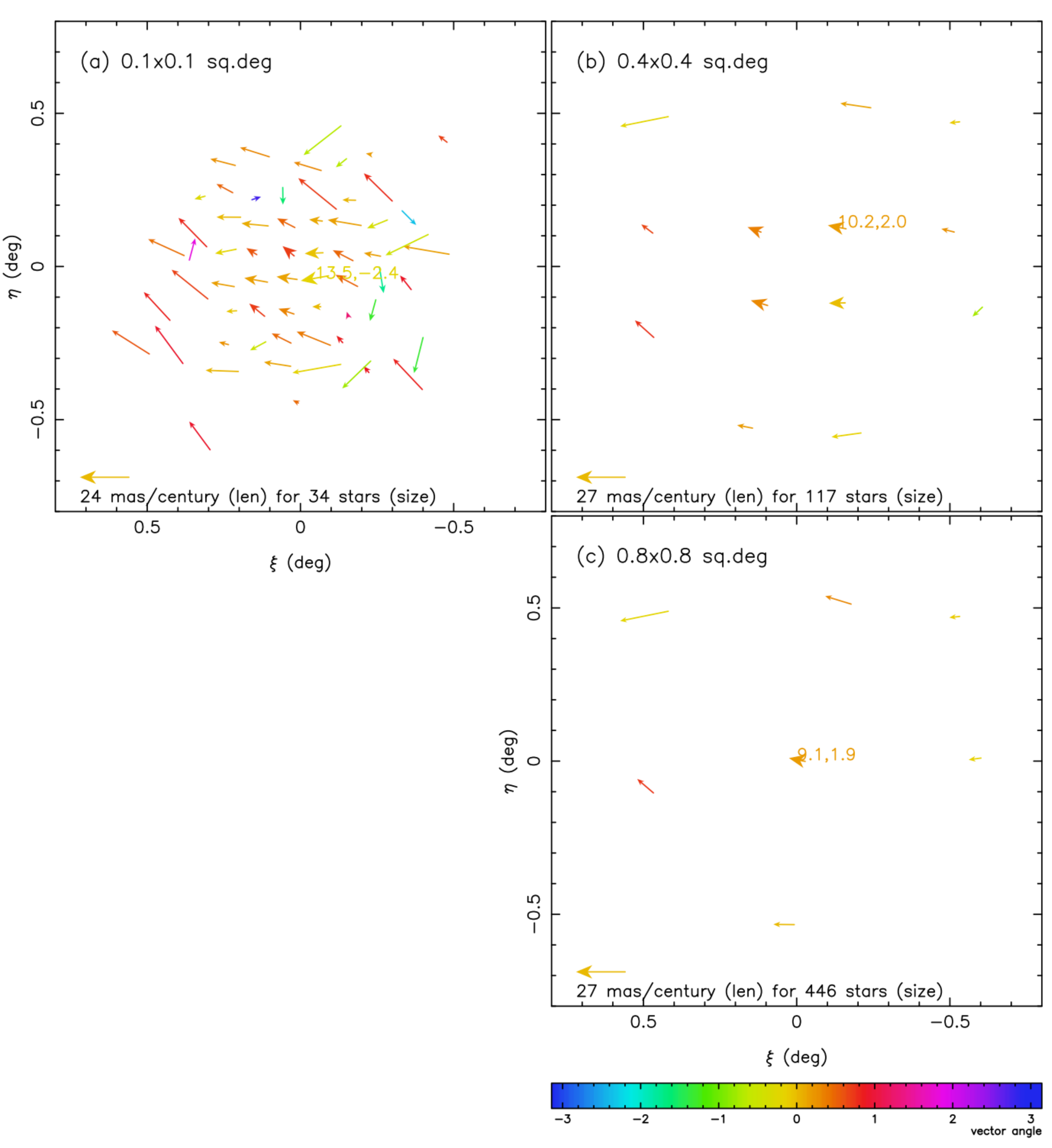}
\caption{Comparison of mean proper-motion maps, plotted in standard coordinates, with $(\xi,\eta) = (0,0)$ corresponding to the centre of the dwarf galaxy. The arrow length is proportional to the proper-motion magnitude, while the arrow-head size is proportional to the number of stars in the bin (scale as indicated in each panel). The direction of the proper-motion vector is given by its colour as well as by the direction of the arrow.  Each realisation (same in each panel) contains $500$ stars, with bin sizes of (a) $0.1\times 0.1 \,\rm{deg}^2$, (b) $0.4\times 0.4 \,\rm{deg}^2$ and (c) $0.8\times 0.8 \,\rm{deg}^2$. Note that each arrow is plotted such that its centre is located at the mean coordinates of the stars in that bin. The bins themselves are set up using a regular grid in $(\xi,\eta)$.}
\label{fig:fig}
\end{figure}

Our aim is to set up a model of Sculptor that is moving through the Galaxy's dark-matter halo in some way, which we then `observe' as Gaia would. In so doing, we test whether it is possible to recover Sculptor's proper motion regardless of the significant measurement uncertainties resulting from its large distance from us. We use the `mock' Sculptor models set up by Breddels {\etalsp} (\cite{2013MNRAS.433.3173B}, Appendix~A), in which a numerical approximation to the distribution function is used to construct the model galaxy from a Plummer light distribution and a NFW dark-matter halo. Each model galaxy consists of $5\times10^4$ stars.

A direct measurement of Sculptor's proper motion\footnote{Proper motions given in the heliocentric rest frame.} exists (Piatek {\etalsp} \cite{2006AJ.131.1445P}; $(\mu_\alpha,\mu_\delta) = (9\pm13,2\pm13)\,\rm{mas/century}$). While an indirect proper-motion determination from stellar redshifts (Walker {\etalsp} \cite{2008ApJ...688L..75W}; $(\mu_\alpha,\mu_\delta) = (-40\pm29,-69\pm47)\,\rm{mas/century}$) shows disagreement with the former measurements, the latter require Sculptor to be non-rotating, while Battaglia {\etalsp} (\cite{2008ApJ...681L..13B}) have shown the presence of statistically significant rotation in the system. We have opted to use the measurement by Piatek {\etalsp} for our exercise with the following steps:

\begin{enumerate}
\item{place the model galaxy `on the sky': give the model system the coordinates and velocity of Sculptor;}
\item{choose a stellar sample: randomly choose $N$ stars out of the mock galaxy catalogue to `observe';}
\item{simulate observational measurements: convolve the proper-motion values of each star ($\mu_\alpha$ and $\mu_\delta$) with a Gaia measurement uncertainty (\SI{147}{\micro as/yr});}
\item{create mean proper-motion map: bin the data into regions of size $x\times x \,\rm{deg}^2$;}
\item{Monte-Carlo simulation: repeat $n$ times, creating a suite of realisations.}
\end{enumerate}

Some of the steps above contain variable parameters. By keeping some constant and varying others, we can explore the effect of changing the parameters on the recovery of the proper motion in a given region of the projected system. Figure \ref{fig:fig} shows examples of the resulting mean proper-motion maps.

The horizontal branch of Sculptor lies at $V\sim20\mag$, with the tip of the red giant branch at $V\sim17\mag$ (Battaglia \cite{Battaglia_thesis}). For our exercise, we choose to use the same uncertainty across all of the stars in the model that are `observed' {\it{i.e.}} each true proper-motion value becomes a Gaussian (whose standard deviation is given by this uncertainty value), from which the `measured' value is randomly chosen. As Gaia will measure proper motions down to $G\sim20\mag$, we take a conservative estimate for the uncertainty --- a fixed value of \SI{147}{\micro as/yr} --- corresponding to the current estimate of the end-of-mission proper-motion uncertainty at $G\sim V\sim19$, one magnitude above Sculptor's horizontal branch.

\section{Conclusions and future work}

Our aim with this work has been to show whether or not Gaia measurements of the proper motions of individual stars within Milky Way dwarf galaxies, whose individual uncertainties could easily be of the order of (or larger than) their true values, could still be used to determine the on-sky motion of the satellite galaxy itself. Our exercise using Sculptor shows that Gaia will measure proper motions of a sufficiently high number of stars to enable the determination of at least the global, systemic proper motion by `beating down' the noise. A qualitative analysis of the proper-motion maps from sets of Monte-Carlo simulations (not presented here), in which we changed the measurement uncertainty to unrealistically optimistic values, appears to show that distinguishing between systems with different velocity anisotropies at such distances would not be easy. How challenging this will be with more realistic measurement uncertainties remains to be quantitatively verified.

The near-polar Galactic latitude of Sculptor ensures that its members will constitute a very large fraction of stars observed by Gaia in its direction. For this reason, an explicit computation of the density of foreground contaminants from the Galactic disk and stellar halo was not performed. A repeat of this exercise for other dwarf galaxies would, however, likely benefit from the use of a mock Gaia catalogue that provides a simulated expectation of the on-sky density of disk and halo stars in the direction of the dwarf galaxy under study. This is particularly important for low-latitude  galaxies, for which contamination of their fields from foreground disk stars cannot be neglected.

Another way in which this test could be improved would be to implement magnitude-dependent Gaia errors by using the galaxy's luminosity function, so that we free ourselves from the use of a constant and conservative estimate of the proper-motion uncertainty, irrespective of the star's magnitude. If this improves the recovery of the global proper motion, it may be worthwhile testing to see whether rotation signals could be retrieved. Finally, it would be informative to investigate how small the proper motion could be for dwarf galaxies in general (in a distance-dependent manner), while still being detectable with Gaia-viable stellar counts.

Gaia's magnitude threshold for radial velocity measurements ($G\sim16\,\mathrm{mag}$) is too bright for stars in most Milky Way dwarf galaxies. However, many radial velocity measurements of individual stars within the galaxies are already available from dedicated programs, and data continue to be acquired through individual surveys of these systems. The promise of the addition of proper motion measurements to these data bodes well for studies of their dynamics under the gravitational influence of the Galaxy's dark-matter halo and, additionally, for studies of these systems as individual entities, also evolving within their own independent dark-matter sub-halos.


\begin{thebibliography}{99}
\bibitem[2007]{Battaglia_thesis} Battaglia, G. 2007, PhD thesis
\bibitem[2008]{2008ApJ...681L..13B} Battaglia, G., Helmi, A., Tolstoy, E., Irwin, M., Hill, V. and Jablonka, P. 2008, ApJL 681, L13

\bibitem[2013]{2013MNRAS.433.3173B} Breddels, M. A., Helmi, A., van den Bosch, R.~C.~E., van de Ven, G. and Battaglia, G. 2013, MNRAS 433, 3173

\bibitem[1998]{1998ARA&A..36..435M} Mateo, M.~L. 1998, ARA\&A 36, 435

\bibitem[2006]{2006AJ.131.1445P}  Piatek, S., Pryor, C., Bristow, P., Olszewski, E.~W., Harris, H.~C., Mateo, M., Minniti, D. and Tinney, C.~G. 2006, AJ 131, 1445

\bibitem[2008]{2008ApJ...688L..75W}  Walker, M.~G., Mateo, M. and Olszewski, E.~W. 2008, ApJL 688, L75

\end{thebibliography}
\end{document}